\documentclass[
prb,twocolumn,10pt,
superscriptaddress,
 amssymb,amsmath,
 aps,
]{revtex4-1}
\usepackage[utf8]{inputenc}
\usepackage{enumitem}
\usepackage[T1]{fontenc}
\usepackage{mathtools}
\usepackage{subfigure}
\usepackage{graphicx}
\usepackage{dcolumn}
\usepackage{bm}
\usepackage{xcolor}
\usepackage{comment} 
\usepackage[normalem]{ulem}

\usepackage{hyperref}
\usepackage{todonotes}
\usepackage{bookmark}

\begin{document}
\author{Ruben Khachaturyan}
\affiliation{Interdisciplinary Centre for Advanced Materials Simulation (ICAMS), Ruhr-University Bochum, Universit\"atsstr 150, 44801 Bochum, Germany}
\author{Yijing Yang}
\affiliation{Interdisciplinary Centre for Advanced Materials Simulation (ICAMS), Ruhr-University Bochum, Universit\"atsstr 150, 44801 Bochum, Germany}
\affiliation{Université Paris-Saclay, Univ Evry, CY Cergy Paris Université, CNRS, LAMBE, 91025, Evry-Courcouronnes, France}
\author{Sheng-Han Teng}
\affiliation{Interdisciplinary Centre for Advanced Materials Simulation (ICAMS), Ruhr-University Bochum, Universit\"atsstr 150, 44801 Bochum, Germany}
\affiliation{Center for Interface-Dominated High Performance Materials (ZGH), Ruhr-University Bochum, Universit\"atsstr 150, 44801 Bochum, Germany}
\author{Benjamin Udofia}
\affiliation{Interdisciplinary Centre for Advanced Materials Simulation (ICAMS), Ruhr-University Bochum, Universit\"atsstr 150, 44801 Bochum, Germany}
\author{Markus Stricker}
\affiliation{Interdisciplinary Centre for Advanced Materials Simulation (ICAMS), Ruhr-University Bochum, Universit\"atsstr 150, 44801 Bochum, Germany}
\author{Anna Gr\"unebohm}
\affiliation{Interdisciplinary Centre for Advanced Materials Simulation (ICAMS), Ruhr-University Bochum, Universit\"atsstr 150, 44801 Bochum, Germany}
\affiliation{Center for Interface-Dominated High Performance Materials (ZGH), Ruhr-University Bochum, Universit\"atsstr 150, 44801 Bochum, Germany}

\title{Microscopic insights on field induced switching and domain wall motion in orthorhombic ferroelectrics}
\email{anna.gruenebohm@rub.de}

\begin{abstract}
 Surprisingly little is known about the microscopic processes that govern  ferroelectric switching in orthorhombic ferroelectrics. 
  To study microscopic switching processes we combine ab initio-based molecular dynamics simulations and data science on the prototypical material BaTiO$_3$. We reveal two different field regimes: For moderate field strengths, the switching is dominated by domain wall motion while a fast bulk-like switching can be induced for large fields. Switching in both field regimes  follows a multi-step process via  polarization directions perpendicular to the applied field. In the former case, the moving wall is of Bloch character and hosts dipole vortices due to nucleation, growth, and crossing of two dimensional 90$^{\circ}$ domains. In the second case, the local polarization shows a continuous correlated rotation via a an intermediate tetragonal multidomain state.

\end{abstract}

\pacs{}

\maketitle

\section{Introduction}
\begin{figure*}
    \centerline{
        \includegraphics[width=1\textwidth, clip, trim=0cm 0cm 0cm 0cm]{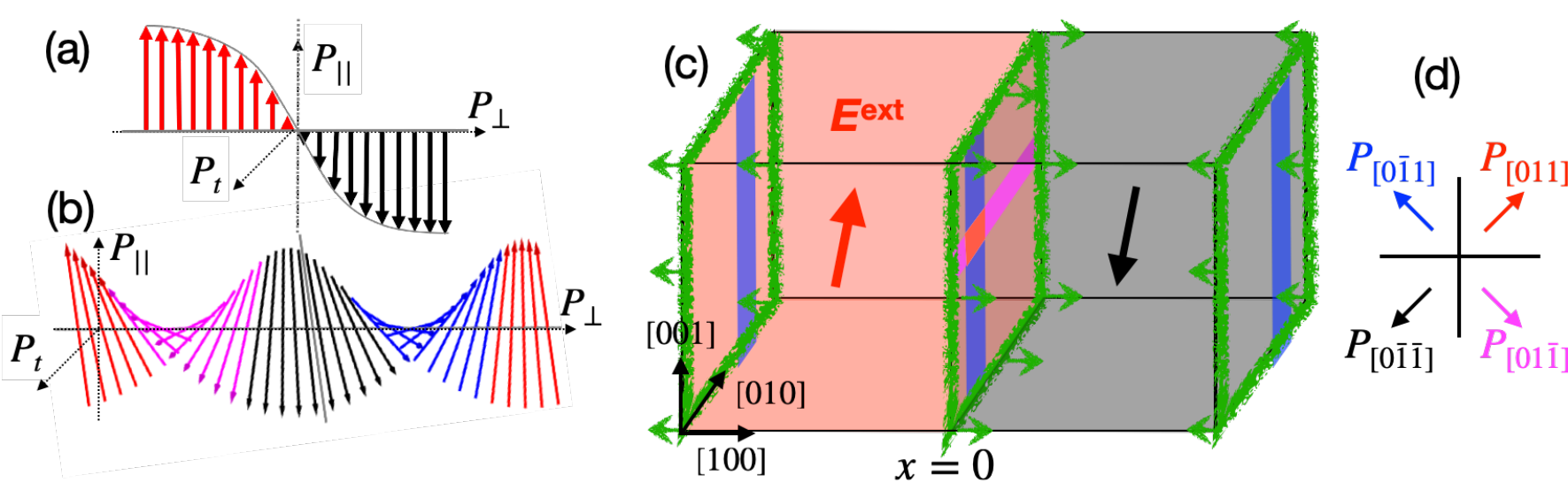}
        }
    \caption{ (a)--(b) Idealized Ising and Bloch walls with 180$^{\circ}$ switching of the polarization via (a) $|P|=0$ ($P_{\perp}=P_t=0$) and (b) the rotation of the polarization via $\pm P_{t}$ ($P_{\perp}=0$, $|P|=const.$).
        (c) Illustration of the simulation setup.
Initially, the system is in the orthorhombic phase with 180$^{\circ}$ domain walls normal to [100] (green frames), 
     a simulation cell of $144\times48\times48$~f.u.\ in combination with periodic boundary conditions is used and the discussion focuses on the wall initially centered at $x=0$. 
    The local polarization in both domains points along $P_{||}=\pm P_{[011]}$ while the components $P_{\perp}=\pm P_{[100]}$ and $P_t=\pm P_{[01\bar{1}]}$ are zero.
(d) Used color encoding:
    We classify the dipoles by the signs of their [011]  components as: $black$:$--$; $red$: $++$, $blue$: $-+$, $magenta$: $+-$.  
   After a field is applied along [011] ($\bm{E}^{\text{ext}}$), the parallel domain (red) grows and the walls move (green arrows). 
  }
    \label{fig:setup}
\end{figure*}
It is an open question whether the reversal of polarization in ferroelectrics happens mainly by direct 
switching (by 180$^{\circ}$), or in  successive multistep events, e.g.\ two 90$^{\circ}$ steps. 
In macroscopic models,  the energy barrier for direct switching is larger than for 
successive switching events.~\cite{nakayama_reconstruction_2014, daniels_Two-step_2014, genenko_stochastic_2018, genenko_multistep_2020,Huang_behaviour_2014} 
The latter has also been reported for single crystals~\cite{Jiang_Direct_2008,Chun_Direct_2011,Dapeng_Dorect_2021} and polycrystalline samples.~\cite{Fancher_X-ray_2017, Schultheiss_Revealing_2018} Furthermore it has been discussed that on the macroscopic scale 
correlated and coherent multistep switching for example by pre-existing  domain structures, cannot be distinguished from direct switching.~\cite{daniels_Two-step_2014, genenko_stochastic_2018} 
 However it is also known that  homogeneous switching is  only relevant for high field strengths and defect free materials.\cite{boddu_molecular_2017} Otherwise, switching is related to nucleation at interfaces or defects and the consecutive growth of nuclei.~\cite{merz_domain_1954, jiang_resolving_2009,grunebohm_interplay_2021,shin_nucleation_2007,klomp_switching_2022,ya_shur_micro_2014} 
The realization of multi-step switching by nucleation and growth is thus challenged by electrostatic and elastic interface energies.

Ferroelectric domain walls (DW) are the interfaces between regions with different orientations of the polarization ($\bm{P}$). 
It is settled that  the characteristics of ferroelectric switching and polarization kinetics depend on domain walls and their dynamics.
However, the underlying microscopic processes are not fully clear and even the nature of the polarization change across domain walls is again under debate.\cite{schultheis_ferroelectric_2023,grunebohm_interplay_2021} 
On charge neutral walls,  the component of the polarization vector parallel to  the wall ($P_{||}$)  switches its sign. For a long time, it was accepted that this switching is Ising-like with  
a local reduction of the magnitude of $\bm{P}$, see Fig.~\ref{fig:setup}~(a).~\cite{nataf_domain-wall_2020,yudin_impact_2012}  However, phenomenological models have predicted that (meta-) stable Bloch-like walls with the rotation of the order parameter, i.e.\ a finite $P_t$ on the wall, see Fig.~\ref{fig:setup}~(b), may exist in orthorhombic~\cite{huang_theory_1997,marton_domain_2010} and rhombohedral ferroelectric phases.~\cite{marton_domain_2010,stepkova_ising_2015,hlinka_phasefield_2011,stepkova_stress-induced_2012,taherinejad_bloch-type_2012} Bloch walls with locally varying rotation direction and Ne\'el like walls with finite polarization perpendicular to the wall have been reported.~\cite{stepkova_ising_2015,cherifi-hertel_shedding_2021} For tetragonal ferroelectrics, Bloch-like walls have been discussed for 
PbTiO$_3$~\cite{wang_origin_2014,wojdel_ferroelectric_2014} and for particular wall directions in Pb-free materials.\cite{yudin_bichiral_2012,gu_flexoelectricity_2014,li_first-principles_2014}
As reviewed by Cherifi-Hertel~\cite{cherifi-hertel_shedding_2021}
there is experimental evidence on non-Ising walls. However Bloch-walls are less common than in theoretical predictions which may be partly related to the surface-sensitivity of the experimental analysis\cite{Zavaliche_Polarization_2005,Salia_Non-Ising_2017,Balke_Topological_2012,hong_vortex_2021} On the other hand, most theoretical predictions are sensitive to the chosen parametrization and may overestimate the polarization rotation. 
 
Microscopic theoretical studies on ferroelectric switching and domain wall motion allow for the fundamental understanding of the underlying nucleation processes and have been successfully used to predict the acceleration of the switching by non-equilibrium effects ~\cite{boddu_molecular_2017,shin_nucleation_2007,khachaturyan_domain_2022,Sepliarsky_polarization_2001,klomp_switching_2022} as well as complex multi-step switching by twinning for 90$^{\circ}$ walls.\cite{xu_ferroelectric_2015} 
However literature mostly focus on tetragonal ferroelectrics, while also orthorhombic phases are common around room temperature, e.g.\ in (K, Na)NbO$_3$ and solid solutions based on BaTiO$_3$.  In one of the few microscopic studies macroscopic multi-step switching has been reported for BaTiO$_3$ in non-collinear electrical fields.\cite{paul_polarization_2009}

In this paper we combine  microscopic simulations  with methods from data science  to 
study field-induced polarization switching and domain wall dynamics of  orthorhombic phase of BaTiO$_3$. We find two switching scenarios: 
For moderate fields, the polarization switches by field-induced motion of domain walls via two dimensional nucleation and growth. Excitingly, this is related to a multi-step nucleation and growth process, Bloch-like polarization rotation and polarization vortices. For high fields, the polarization switches via an ultrafast homogeneous switching path which sets in in the center of the domain and propagates in all directions in 3-dimensional space.

\section{Methods}
We use molecular dynamics (MD) to simulate the field-driven  evolution of the  polarization vector field. Concretely, we use the effective Hamiltonian by Zhong {\it{et al.}}~\cite{zhong_first-principles_1995} parametrized by \textit{ab initio} simulations~\cite{nishimatsu_first-principles_2010} and the \textit{feram} code.\cite{nishimatsu_fast_2008} 
This approach allows to efficiently model the microscopic properties of ferroelectrics as the collective atomic displacements are coarse-grained to the most relevant degrees of freedom per unit cell: the acoustic displacement vector $\bm{w}$ and the soft mode vector $\bm{u}$, which corresponds to the local dipole moment $\bm{p} = Z^*\bm{u}$, with $Z^*$ the effective Born charge.
The Hamiltonian is given as
\begin{eqnarray}
      \label{eq:eff_Hamiltonian}
      \nonumber H^{\rm eff}
     & =& V^{\rm self}(\{\bm{u}\})+V^{\rm dpl}(\{\bm{u}\})+V^{\rm short}(\{\bm{u}\})\\
      \nonumber &+& V^{\rm elas}(\eta_1,\dots\!,\eta_6,\{\bm{w}\}))\\
      \nonumber &+& V^{\rm coup}(\{\bm{u}\},\{\bm{w}\}, \eta_1,\cdots\!,\eta_6)\\
      &-&Z^*\sum_i \bm{E}^\text{ext} \cdot \bm{u}_i  + \frac{M^*_{\rm dipole}}{2} \sum_{i,\alpha}\dot{\bm{u}}_{\alpha,i}^2,
    \end{eqnarray}
where $V^{\rm self}(\{\bm{u}\})$, $V^{\rm dpl}(\{\bm{u}\})$ and $V^{\rm short}(\{\bm{u}\})$ are the self-energy, the long range dipole-dipole interaction, and the short range interactions of the local soft modes, the elastic energy $V^{\rm elas}$ depends on $\{\bm{w}\}$ and the homogeneous strain tensor $\eta_i$ given in Voigt notation, and $V^{\rm coup}$ includes the couplings between local soft mode and strain.
The last two terms are the coupling to the external field $\bm{E}^{\text{ext}}$ and the kinetic energy of the local mode, with  $M^*_{\rm dipole}$ being the effective mass of the soft mode.
The strain is internally optimized during each MD step. 
Previous work demonstrated the high predictive power of this method for ferroelectric domain walls and their dynamics.~\cite{lai_domain_2007,grunebohm_ab_2015,khachaturyan_domain_2022}
 
Figure~\ref{fig:setup}~(c) sketches our simulation set up. Without loss of generality, we study the orthorhombic phase with polarization along $\pm[011]$ and an external field ($\bm{E}^{\text{ext}}$) applied along $[011]$ and focus on typical 180$^{\circ}$ domain walls with $\langle 100\rangle$ normal (O180), \cite{marton_domain_2010,soergel_visualization_2005} see Fig.~\ref{fig:setup}.
We use a system size of $144\times48\times48$ f.u.\ together with periodic boundary conditions and initialize a periodic array of 180$^\circ$ DWs in $x$-planes. 
For the DW distances of 28~nm, interactions between neighboring walls are negligible.\cite{grunebohm_domain_2012,klomp_switching_2022}
Furthermore, finite-size effects and thermal noise are small and we validate our results by independent simulation with different random initialization of dipoles.

We apply the field protocol from Ref.~\onlinecite{khachaturyan_domain_2022} to equilibrate 
the system in this well-defined domain structure, i.e.\ we randomly initalize local dipoles while applying local external fields and thermalize the system in the orthorhombic phase (40\,K apart from the T-O transition temperatures found for cooling  by our model)  using the Nos$\acute{\text{e}}$-Poincar$\acute{\text{e}}$ thermostat.~\cite{bond_nosepoincare_1999}
The local fields are removed and after further equilibration, $\bm{E}^{\text{ext}}$ is applied instantaneously, and we analyze the time evolution of the  polarization vector field  on slices parallel to the wall ($\langle p_z\rangle_x$), i.e.\ on  layers $x$, see Fig.~\ref{fig:setup}~(c). \footnote{Note that the multistep switching on the moving walls was also reproduced applying the field with a ramping rate 1~kV/mm per picosecond.}
We track the  evolution of the domain wall width ($d_{DW}$) and the velocity of the domain wall center ($v_{\text DW}=\dot{x}_0$)
 by fitting the mean polarization per layer at each time step via 
    \begin{equation}
        \langle p_z\rangle_x = p_0\cdot\tanh{\left[\frac{x-x_0}{d_{DW}}\right]} + \varepsilon(x)\cite{khachaturyan_domain_2022}
        \label{eq:tanh_DW}
    \end{equation}
where the fitting parameters $p_0$ and $\epsilon$ correspond to the saturation polarization without an external field and time and thus space-dependent corrections, respectively.\footnote{Note that the converse piezoelectric effect results in a time-dependent increase of polarization with domain wall motion.}
Complementary, we estimate $v_{DW}$ in the steady state  based on the macroscopic change of polarization $\Delta P$ between $30$~ps and $100$~ps
\begin{equation}
    v_{DW}=\frac{a_0*L_x}{4}  \frac{\Delta P(E)}{t} \frac{1}{P_{DW}}, \label{eq:v}
\end{equation}
where $P_{DW}$, $a_0$ and $L_x$ are the polarization gain by switching one layer, the lattice constant and the number of unit cells along $x$-direction, respectively.
We analyze the formation of polarization vortices on the moving wall by integrating the  winding angle $\Delta\phi_i=\arccos{\frac{\bm{u_{i+1}}\cdot\bm{u_i}}{|\bm{u_{i+1}}||\bm{u_{i}}|}}$ enclosed between  neighboring dipoles along the closed loop in counterclockwise direction on all lattice positions 
\begin{equation*}
        \Phi_i=\frac{1}{2\pi}\sum_{i=1}^{4}{\Delta\phi_i}.\cite{tricocheContinuousTopologySimplification2001}
    \end{equation*}

We furthermore analyze the time-evolution of the distribution of local dipoles $p_i$ and clusters of dipoles. This analysis is done on separate $x$-planes (slices).
Therefore, we classify the local dipoles by the sign of their [011] components, see Fig.~\ref{fig:setup}.: 
\begin{itemize}[leftmargin=1.5cm]
\item[$red$:] $p_z>0 \wedge p_y>0$ ($P_{||}$, parallel to $\bm{E}^{\text{ext}}$),
\item[$black$:]  $p_z<0 \wedge p_y<0$ ($P_{||}$, anti-parallel to $\bm{E}^{\text{ext}}$)
\item[$magenta$:] $p_z>0 \wedge p_y<0$ ($P_t$, perpendicular to $\bm{E}^{\text{ext}}$)
 \item[$blue$:]  $p_z<0 \wedge p_y>0$ ($P_t$,  perpendicular to $\bm{E}^{\text{ext}}$).
 \end{itemize}
\label{desc:critical_clusters}
and define a cluster as a connected region of unit cells (cross-shaped stencil) within the same class on one plane. 
The presented statistics and averages are taken across the evolution of all clusters and times.
\footnote{
 Diagonally-connected cells are not counted as neighbors.} 
The cluster analysis is realized through a wrapper around the labeling capabilities as implemented in the \texttt{skicit-image} library.~\cite{Walt2014}

Note that thermal fluctuations cannot be clearly separated from initial nuclei formation.
To exclude the former, we start with ten independent simulations per field strength over the time interval 3--100~ps and filter out all cluster sizes which exist for one time step only (1~ps).
We further exclude the final cluster, i.e.\ the fully switched slice.
Our dataset now consists of the time evolution of all slices of all ten simulations.
Based on cluster area histograms of vanishing clusters, we extract a critical area by fitting an exponential.
The critical area is then extracted as the intersection of this exponential with occurrence frequency ``1''.
This is what we define as the critical area because it is the upper bound for unstable clusters.
At the same time, this value is a good approximation of the lower bound for the critical nucleus area or extent.
For all clusters, which grow at least as large as these values, we subsequently assess the time-evolution of their extent and area at the critical cluster size using a quadratic fit. 
All scripts used for the analysis are available on our git repository.\cite{git}
\section{Results}
\begin{figure}
    \includegraphics[width=0.45\textwidth]{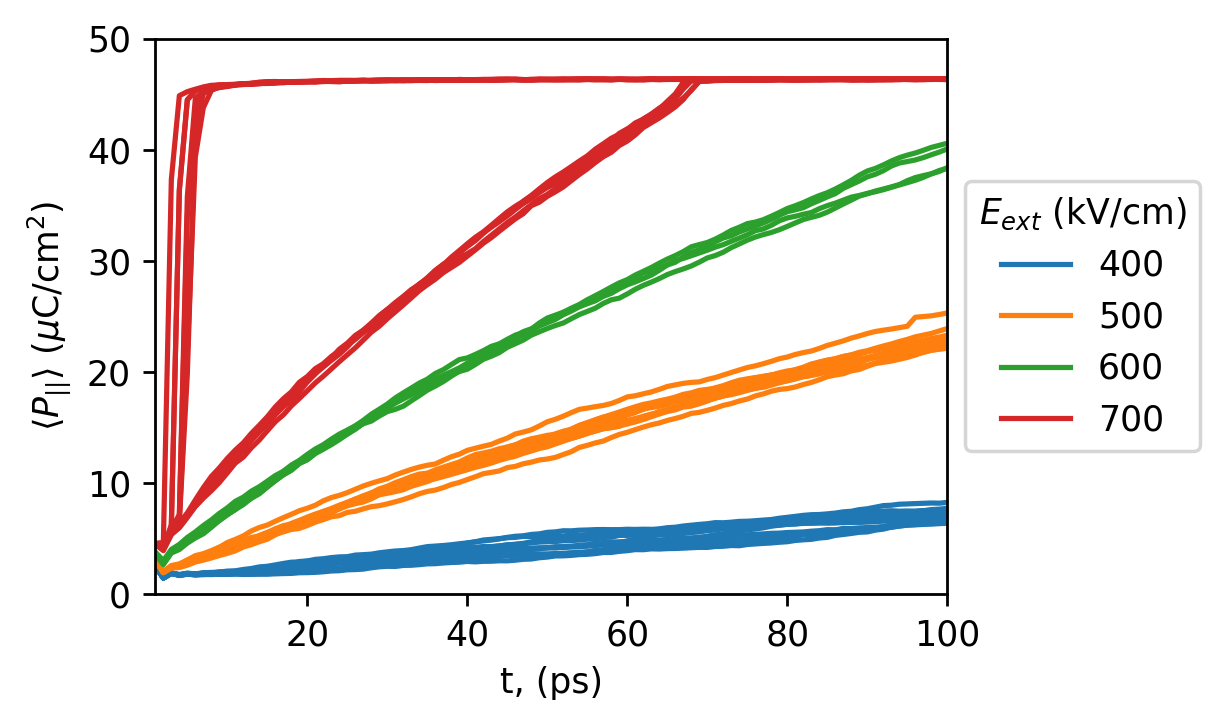}
    \caption{Time-evolution of the macroscopic polarization component $P_{||}$ after external fields of blue: 400~kV/cm, orange: 500~kV/cm, green: 600~kV/cm and red: 700~kV/cm have, been applied to independent initial states. Initially the system is in the   bi-domain state with equal domain width ($P_{||}=0$). For 700~kV/cm about 8 of 10 simulations follow the fast switching scenario discussed in Sec.~\ref{sec:high}
 \label{fig:velocity}
  }
\end{figure}

We start our analysis with a systematic screening of different field strengths. One can distinguish three  switching regimes, see Fig.~\ref{fig:velocity}:
\begin{itemize}
\item[(I)] Below 400~kV/cm  the change of polarization with time is minor (not shown).
\item[(II)] Above this field strength, the polarization increases with time until the saturation polarization of the single domain state in the external field is reached.
Thereby, the velocity of the switching is initially accelerated, and a steady state with approximate linear $P(t)$ is reached after about 30~ps. The speed of the walls increases with the field strength. This switching by wall motion discussed in Sec.~\ref{sec:DW}.
\item[(III)] For 700~kV/cm the system already switches to the single-domain state before the steady state is reached in about 8 of 10 simulations. This bulk-like switching discussed in Sec.~\ref{sec:high}. 
\end{itemize}
Both, the activation field for wall motion and the increase of the switching velocity with the field strength are well-known for ferroelectric switching.~\cite{schultheis_ferroelectric_2023} Furthermore, the initial acceleration of the walls \cite{khachaturyan_domain_2022} and the bulk-like switching for high fields\cite{boddu_molecular_2017} have been reported for  tetragonal BaTiO$_3$.
However, as discussed in the following the underlying microscopic processes differ considerably between  tegragonal (T) and  orthorhombic (O) phases.

\begin{figure}
  \includegraphics[width=0.35\textwidth,clip,trim=0cm 0cm 0cm 0cm ]{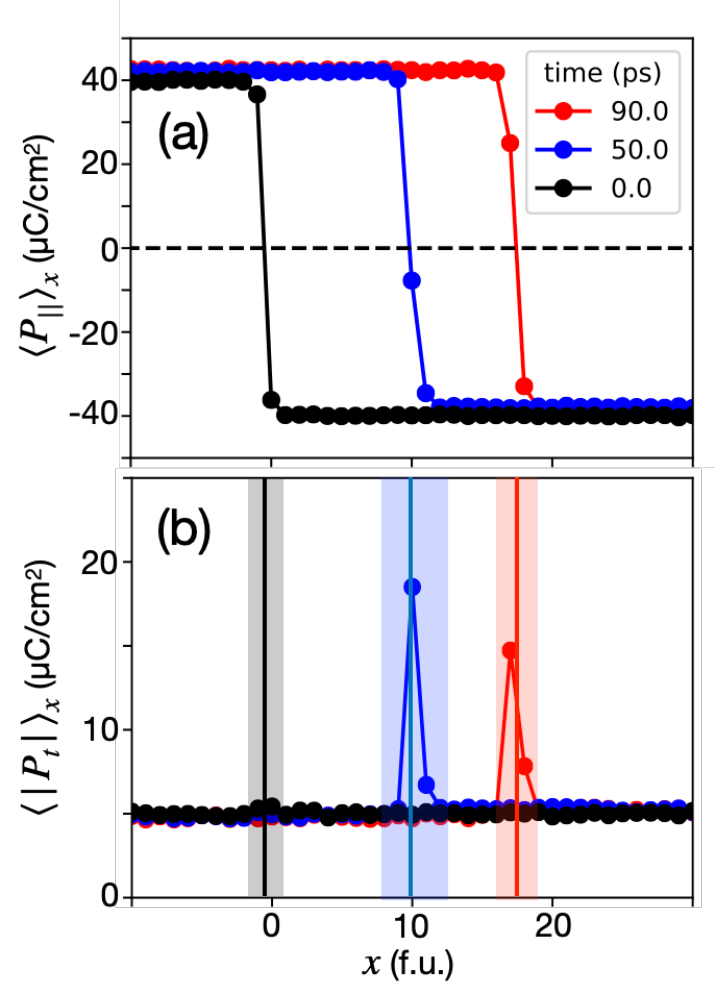} 
    \caption{Polarization profile across an O180 domain wall 
    (0~ps, black) and 50~ps (blue) and 90~ps (red) after an external field of  
   $\bm{E}^{\text{ext}}=500$~kV/cm has been applied. (a) $P_{||}$, i.e.\ the polarization projected on on the field direction [011], (b) $P_{t}$, i.e.\ the polarization  projected on [0$\bar{1}$1].
    Note that for $P_{t}$  the sign of the projection has been removed before averaging. Vertical lines and shaded areas highlight the center and the width of the domain wall at the corresponding time steps. 
 }
    \label{fig:profile}
\end{figure}
\subsection{Field induced domain wall motion}
\label{sec:DW}
What are the underlying microscopic processes in switching regime (II)?
The switching process is dominated by wall motion as the velocity of walls determined by the change of macroscopic polarization, Eq.~\eqref{eq:v}, is close to the actual domain wall speed (Eq.~\eqref{eq:tanh_DW}, added in brackets) with 79.7$\pm$20~m/s ($71$m/s), 128.9$\pm$41~m/s ($119$m/s) and 183.7$\pm$21~m/s ($202$m/s) for 500~kV/cm, 600~kV/cm and 700~kV/cm.
As discussed in literature for the tetragonal phase \cite{khachaturyan_domain_2022,klomp_switching_2022,shin_nucleation_2007} thermal fluctuations of local dipoles increase on domain walls due to frustrated interactions, reduced polarization and shallower energy minima and the  critical field strength for the onset of switching is locally reduced. 

Figure~\ref{fig:profile} shows the polarization profile across the initial domain structure (black) and 50~ps (blue) and 90~ps (red) after the application of 500~kV/cm. 
Initially, the walls are sharp  with a width of about 1~f.u.\, and we can reproduce the prediction by Landau theory~\cite{marton_domain_2010} that the equilibrium wall configuration is of Ising-type, i.e.\  $P_{t}$ and $P_{\perp}$ at the wall can not be distinguished from  thermal noise. In the steady state of the motion, the mean wall width increases to  1.2~f.u.\ ($\pm 10$\%).    
Excitingly, the applied field changes the character of the walls to mixed Ising-Bloch-like with $\langle |P_{t}| \rangle \approx 20$~$\mu$C/cm$^2$, see Fig.~\ref{fig:profile}~(b). On the other hand, no Ne\'el-like polarization rotation is induced as $P_{\perp}$ changes neither globally nor locally, cf. Fig.~\ref{fig:cut_plane}. 

 \begin{figure}
  \includegraphics[width=.45\textwidth,clip,trim=0cm 0cm 0cm  0cm]{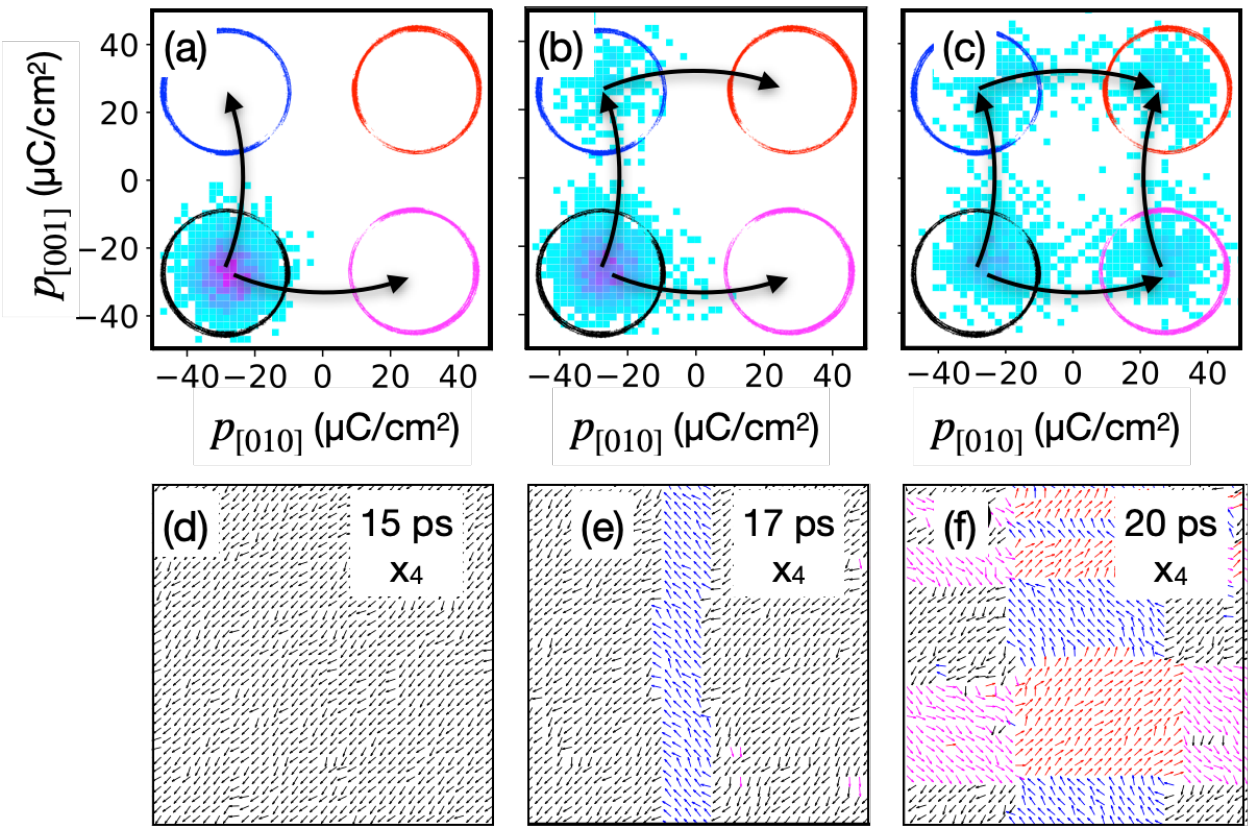}
        \caption{ 
        Time-evolution of the dipole distribution in layer $x_4$ (a)--(c) in $p_{[010]}-p_{[001]}$ space  and (d)--(f) in real space after 
  500~kV/cm  have been applied along the $red$ direction. Colors encode (a)--(c) the number of dipoles per (2.06 $\mu$C/cm$^2$)$^2$ area or (d)--(f) the dipole direction
  and black arrows show the proposed transition path.
  (a)/(d) 15~ps (distance to DW center: 1.6~f.u.), (b)/(f) 17~ps (distance to DW center: 1.1~f.u.), and (c)/(g) 20~ps (distance to DW center: 0.1~f.u.).  
        }
\label{fig:dist}
\end{figure}
Exemplary, Figure~\ref{fig:dist} shows the time-evolution of the dipole distribution on a slice parallel to the domain wall for 500~kV/cm. Initially, the dipoles  are in the metastable $black$ state 
antiparallel to the applied field and all dipoles scatter around the mean value of polarization of about $P_{||}=-37~\mu$C/cm$^2$, see Fig.~\ref{fig:dist}~(a) and (d). 
After full switching (not shown) the dipoles analogously scatter around the mean value of the $red$ state $P_{||}=43~\mu$C/cm$^2$ which is the energetic ground state in the applied field. 
During switching, the dipoles have to cross the energy barriers between these two states. 
As the probability for the realization of a particular point in configuration space depends on its energy,\cite{ocenasek_dynamics_2023} subfigures (a)--(c) allow us to draw important conclusions on the switching process:
On the one hand, no dipoles are present in the center of the $p_{[001]}-p_{[010]}$-plane for all times.
The barrier corresponding to direct $180^{\circ}$ switching is thus too high for the given field strength and temperature. 
 On the other hand,  the $\langle 100\rangle$-type directions are lower in energy
 and are populated at intermediate times.
 For the shown example a fraction of dipoles already crossed one of these barriers at 17~ps entering the $blue$ state, i.e.\ the local energy minimum with $P_{[010]}>0$ and $|P_{[010]}|\approx |P_{[001]}|$ with polarization perpendicular to the applied field. Thus, the two-step $90^{\circ}$ switching across the $\langle 100\rangle$-type directions is not only lower in energy than direct switching in homogeneous phenomenological models~\cite{genenko_multistep_2021} but also on the microscopic scale. 

It is important to note that the macroscopic polarization does not switch in two discrete steps in time. For all times, a variety of intermediate angles of polarization is present.  This multistage switching is activated by thermal fluctuations of local dipoles but does not occur homogeneously in space. 
This can be understood by the the polarization gradient and the bound charges $\rho_i = -\left( \partial P_{i}/\partial i \right)$ related to a depolarization field and an energy penalty proportional to $\rho^2$~\cite{shin_nucleation_2007}
induced if locally the sign of the polarization component $i$ switches relative to its surrounding. 
Therefore, the $z$ and $y$ interfaces of  {\it blue} and {\it magenta}  and all interfaces of {\it red} clusters with the $black$ matrix are high in energy, respectively.
The switching happens by nucleation and growth of local clusters, see subfigures (e)--(f).

\begin{figure}
    \centering
  \includegraphics[width=0.4\textwidth,clip,trim=0cm 0cm 0cm 0.0cm]{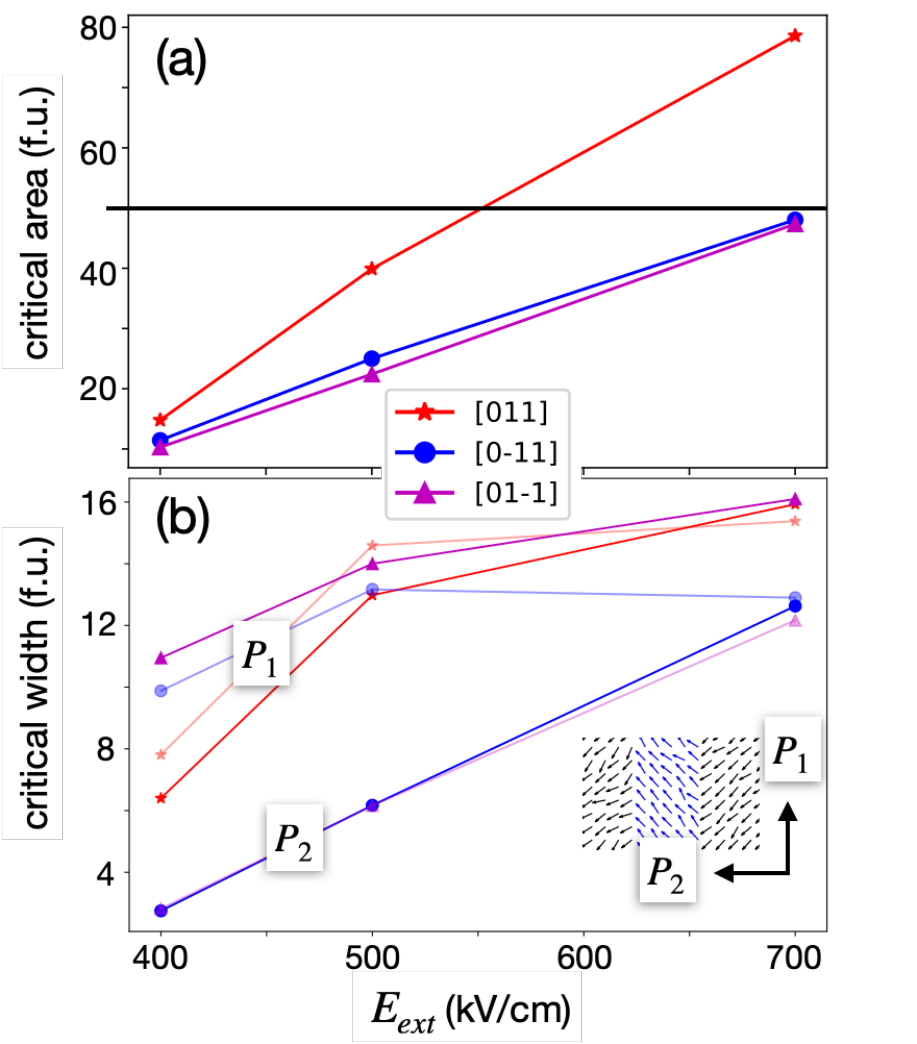}
    \caption{Change of critical cluster (a) area and (b) width for 
 $blue$, $magenta$, and $red$ clusters as a function of field strength. In (b) one can distinguish the critical width for the direction without  ($P_{1}$) and with 
  ($P_{2}$) reversal of the sign of the polarization component 
    as illustrated in the inset for the $blue$ cluster.  
       \label{fig:crit_cluster_area} }
\end{figure}

As discussed in  Sec.~\ref{desc:critical_clusters} initial nucleation and thermal fluctuations can not easily be distinguished and we thus define disappearing and persisting clusters as approximate measures. 
Thereby, it is convenient to distinguish between the initial switching to $blue$ and $magenta$ directions
and the final switching to the $red$ direction.  
For 400~kV/cm the number of disappearing clusters is more than 5 times larger for initial than for final switching while the number of persisting clusters is about 2 times larger for final switching.
 This shows that thermal fluctuations and nucleation are more relevant for initial switching and both switching steps follow different dynamics. 
 Increasing the field strength, the number of initial disappearing clusters is reduced (by a factor of 1.5 for 700~kV/cm) as more clusters form stable nuclei and the number of persisting clusters increases by a factor of about 3.

Figure~\ref{fig:crit_cluster_area}~(a) compares the average critical areas (i.e.\ number of f.u.\ in a cluster) for different field strengths. Note that in agreement to the tetragonal phase,\cite{ocenasek_dynamics_2023} the clusters are not perfectly rectangular. The difference between $blue$ and $magenta$ clusters is the error 
given by the used fitting and the finite number of statistical samples as 
equal properties of $blue$ and $magenta$ clusters are expected in the thermodynamic limit. For 400~kV/cm the critical cluster size for initial switching is approximately 10~f.u. 
With increasing field strength, field-induced fluctuations and this area increase linearly to about 50~f.u. 
For $red$ clusters the critical area is larger and increases linearly to about 80~f.u. This is mainly because much less smaller clusters form by thermal fluctuations while the area of crossing initial clusters switching to $red$ increases with the field strength.

\begin{figure}
    \centering
    \includegraphics[width=0.35\textwidth,clip,trim=1cm .5cm 1cm .5cm]{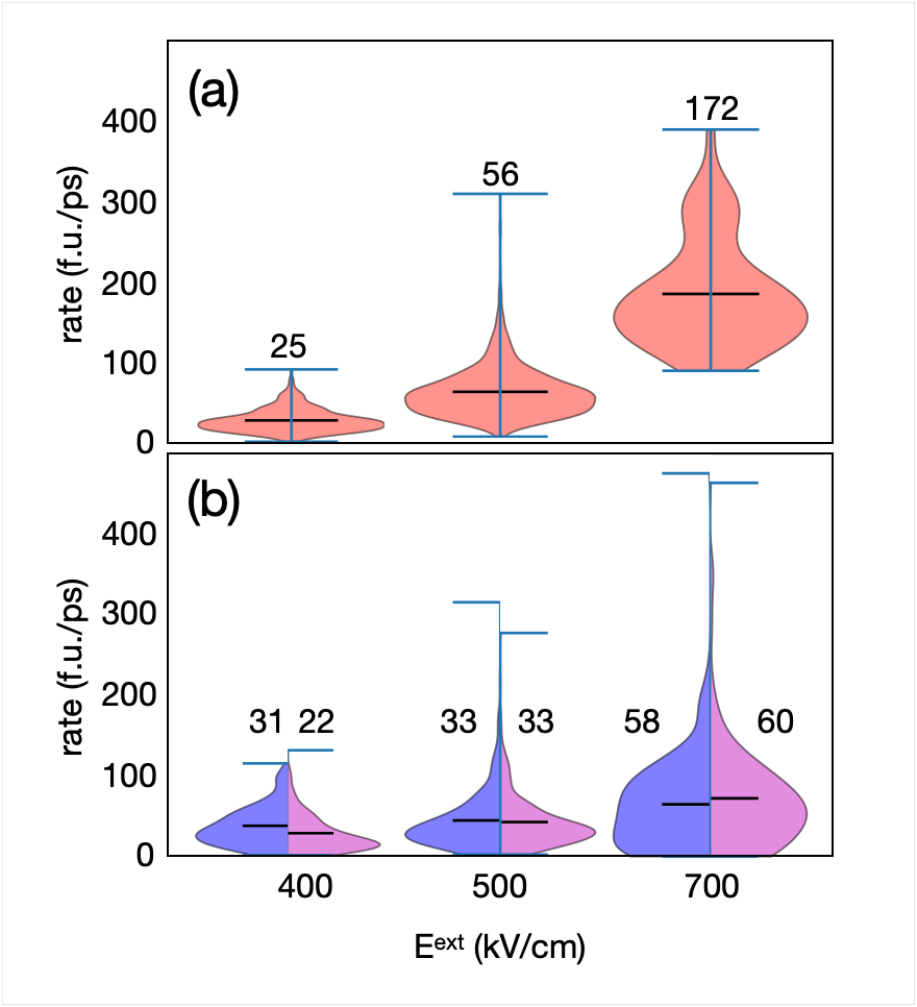}
    \caption{
    Distribution of the growth rate of clusters of the size estimated in the previous figure depending on field strength (a) for $red$ clusters and (b) for $blue$ and $magenta$ clusters.  Black horizontal lines mark mean values and the lower and upper "whiskers" extending from the violin plot indicate the minimum and maximum values of the data distribution, respectively. Numbers are the median of each distribution in f.u./ps. 
    \label{fig:cluster_growth_rates}}
\end{figure}

It is  important to note that due to the bound charge formation on $z$ and $y$ interfaces of $blue$ and $magenta$ clusters, respectively, 
the initial clusters and their growth are not isotropic. 
Figure~\ref{fig:crit_cluster_area}~(b) shows the dependency of the critical extents  on the field strength. 
Persisting initial clusters show critical extents
along the long and short axes in the range of 10~f.u and 3~f.u. For the $red$ cluster the difference of the extent is on the level of noise (below 2~f.u.). 
Once formed, the growth of the clusters depends on the field strength. 
Figure~\ref{fig:cluster_growth_rates} shows the field dependency of the distribution of the averaged growth rates in f.u./ps of clusters at critical area.
For $red$ clusters, the distributions are symmetrically mirrored in violin plots (a), while the difference between $blue$ and $magenta$ clusters is compared in the left and right parts of each violin in (b).
The average median of the initial clusters increases from about 26~f.u./ps at $400\,$kV/cm to about 33~f.u./ps and 59~f.u./ps at 500 and $700\,$kV/cm, respectively. 
Furthermore, increasing the field strength results in the broadening of the distributions, both at the main peak and the tail, cf.\ $blue$ vertical line segments above the violins highlighting each maximum.
The $red$ direction shows similar median values at $400$ and $500\,$kV/cm.
At $700\,$kV/cm its minimal switching rate increases to 90~f.u./ps and a second peak  around 300~f.u./ps appears. This again hints to the fact that the final switching in large fields is dominated by the crossing of large initial clusters which switch to red within few time steps.

\begin{figure}
\includegraphics[width=0.35\textwidth]{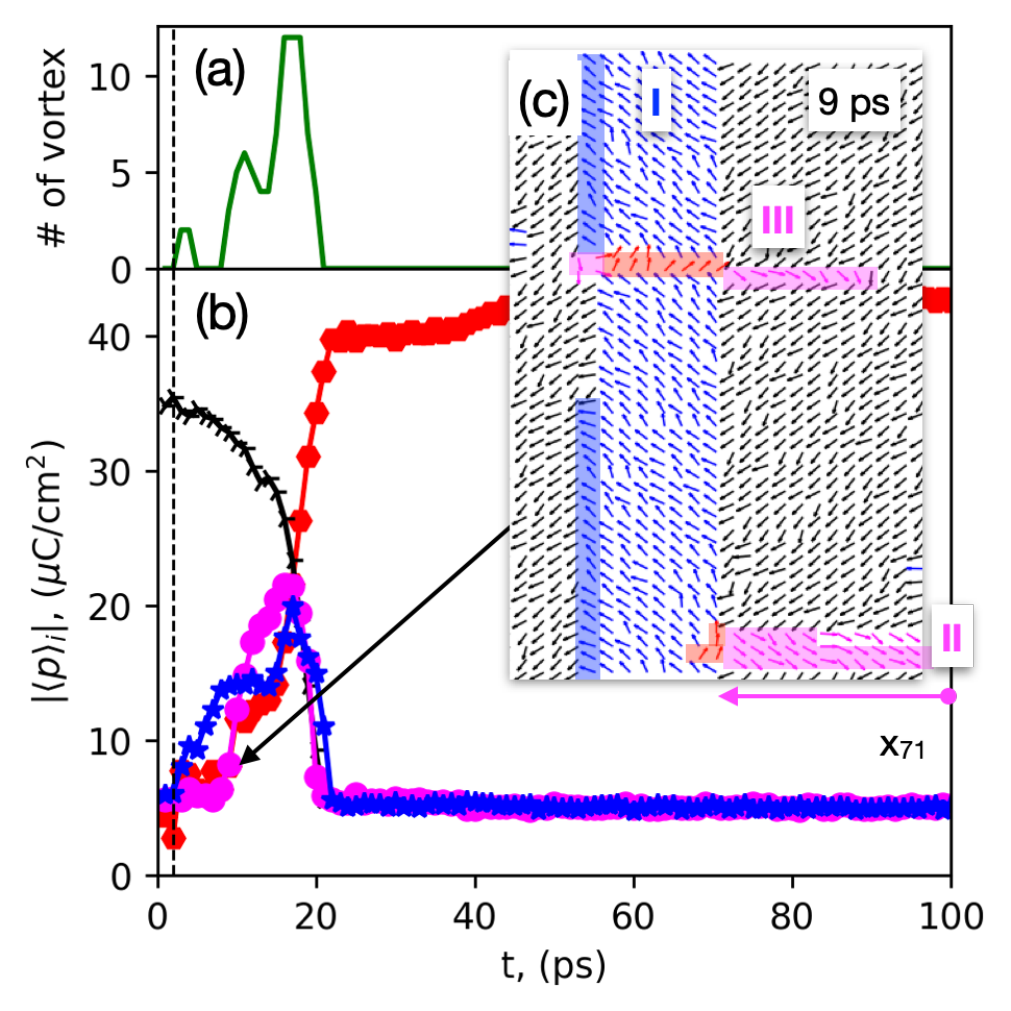}
\caption{Exemplary switching scenario on the moving domain wall for 400~kV/cm and layer $x_{71}$. 
(a) Time-evolution of the number of vortices on that layer. (b) Change of the mean polarization along $red$, $blue$, $magenta$ and $black$ directions. An excerpt of the slice is shown for the time marked by the black arrow. Colors encode the direction of the polarization and the colored background highlights the growth of clusters in the last ps. Three clusters can be distinguished: (I) and (II) $blue$ and $magenta$ clusters nucleated before  (III)  $magenta$ or $red$ cluster nucleated within the last ps. }
  \label{fig:vortex}
\end{figure}

Two typical scenario of nucleation and initial growth are depicted in the inset of Figure~\ref{fig:vortex}: 
The light background colors mark the change of the clusters in the last ps. 
Within this short time  span the $blue$ (I) and $magenta$ (II) clusters have grown by about 40~f.u. and 6~f.u., respectively, and crossed. Thereby forward switching from $blue$ to $red$ sets in at the interface.
One $red$ cluster thus forms by the crossing of two initial clusters which is the most likely scenario for large field strength as discussed above. 
In addition to this scenario, also forward switching by the nucleation of a new $red$ cluster (III) may be possible. However the time resolution is not sufficient to disentangle this process from the nucleation of and crossing of a new $magenta$ cluster (III). 

For both switching scenarios, the mean polarization on the slice follows the trends shown in Figure~\ref{fig:vortex}~(b): 
Either $blue$, $magenta$ or both types of clusters nucleate. 
Due  to the depolarization field these  initial clusters  commonly span the whole system along the depolarization direction within less than 2~ps,\footnote{Verified for 10 checked samples}  i.e.\ with a speed of approximately 4700 m/s close to the velocity of sound in the material for all field strengths.\cite{lee_elastic_2019} To relate this to the switching behavior of a single cluster in a slice this means that a cluster with an initial extent of 3~f.u.\ covers the whole slice length within 2~ps.
After that the different regions on the wall are  separated by one dimensional O90 walls running along [010] and [001] with a typical width of less than 2~f.u. which show a slower  side-wise motion. Thus each wall splits into 2D regions 
with either $black$ and  $red$ interfaces (Ising-type wall) or intermediate $blue$ or $magenta$ polarization (Bloch-type wall). 
In this transient state polarization vortices form at the boundaries between the different type of clusters, see subfigure (a). The number of vortices is always maximal if the domain wall is centered on a particular layer. However as soon as the domain wall passed the slice, or if the field is switched off the vortices disappear.
The final switching is fast as the whole crossings of both domain types switch at once.  Furthermore typically several initial clusters have already nucleated at this stage resulting in faster switching by merging. 
 With increasing field strength, the probability for initial nucleation as well as the for merging and crossing of cluster increases resulting in the discussed increase of the domain wall speed.

\subsection{Generalization to other domain walls}
  \begin{figure}
 \includegraphics[width=0.45\textwidth,clip,trim=0cm 0cm 0cm  0cm]{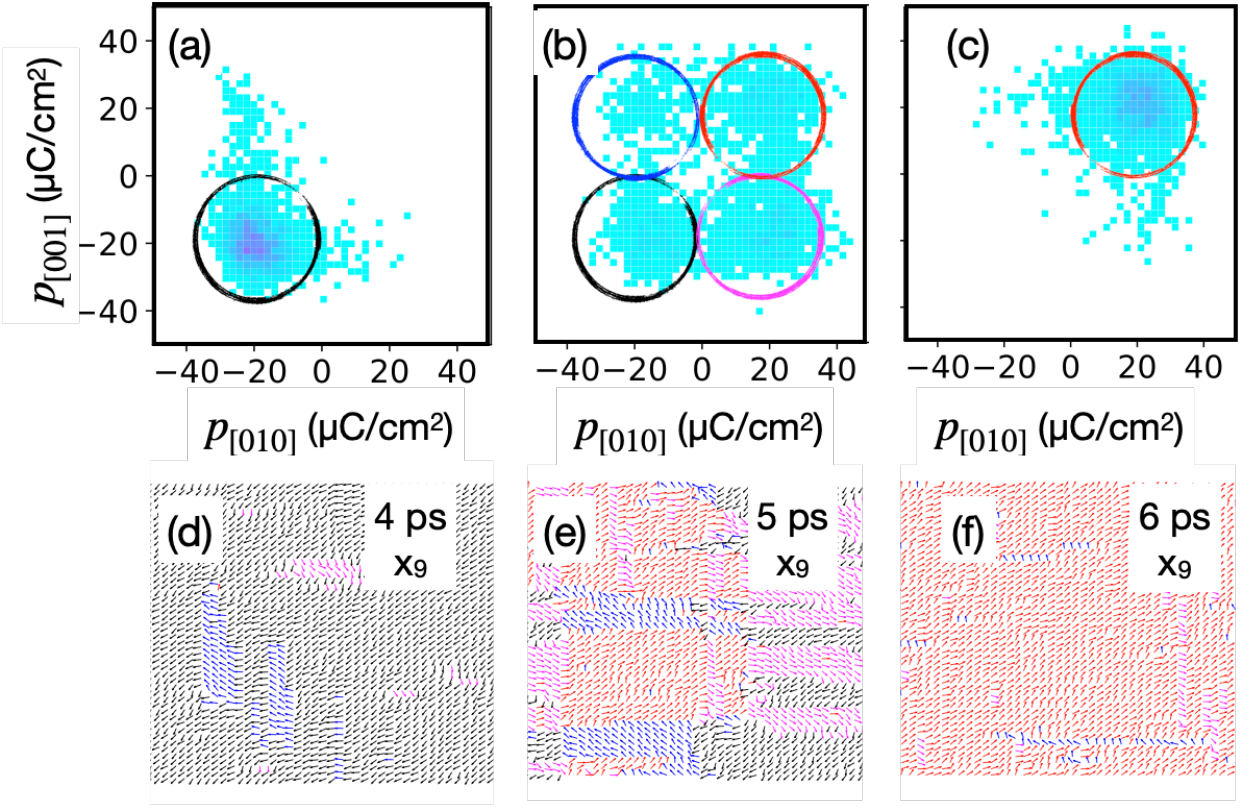}
        \caption{Rhombohedral phase with initial R109 domain walls  at 50~K
                after a field of 300~kV/cm has been applied along [111].
Time-evolution of the dipole distribution in layer $x_9$ (a)--(c) in $p_{[010]}-p_{[001]}$ space  and (d)--(f) in real space. Colors encode (a)--(c) the number of dipoles per (2.06 $\mu$C/cm$^2$)$^2$ area or (d)--(f) the sign of the [010] and [001] components of polarization.
 Note that the distribution of $p_x$ is approximately constant with time. 
        \label{fig:distR}}
    \end{figure}
Why is the discussed 90$^{\circ}$ rotation of the polarization of a fraction of dipoles possible and favorable on the wall? Due to the large electro-mechanical coupling, this observation is unexpected.
What are the properties of the O180 walls allowing for the discussed multi-step switching and is this scenario possible to other types of walls?

First, one may expect that the multi-step switching by nucleation and growth is  favorable, if the transient state is low in energy.  Indeed $blue$ and $magenta$ polarization directions are metastable orthorhombic states and clusters with these polarization direction and the  $black$  initial state share elastically and electrically compatible  O90 walls which are low in energy\cite{marton_domain_2010} as well as stable down to few unit cells.\cite{grunebohm_impact_2020}
Second, the answer might be hidden in the bound charge that forms if a local dipole switches the sign of polarization component $i$. In the case of 90$^{\circ}$ switching all dipoles changes only one polarization component, reducing the bound charge by a factor of about two compared to the full  180$^{\circ}$ switching.
Third, one may expect a larger probability for nucleation by thermal fluctuations, if either one polarization component switches or two highly correlated components switch at the same time. On the other hand, the instantaneous switching of two uncorrelated components is rather unlikely. As discussed above, only one component switches on the moving O180 wall.
Forth, the energy barrier for direct switching needs to be higher than the energy barriers for the multiple switching events. As the free energy of the O phase is closer to the T phase (barriers for the switching of one component) than to the C-phase (barrier for switching via zero polarization) this is indeed true for the O180 wall.

To test these hypotheses, we compare our results for the O180 wall to one example without multistep switching (T180) and one example with multistep switching (R109) walls. 
One the one hand, multistep switching by 90$^{\circ}$ would be possible on the moving T180 wall via a transient T90  state. However, for such switching two polarization components had to switch at the same time and these walls would induce an elastic distortion, i.e.\ are higher in energy and cover a larger volume. 
Furthermore, the difference of free energies between paraelectric and tetragonal phases is smaller than between paraelectric and orthorhombic phases and thus also the energy barrier for direct Ising type switching is probably smaller compared to the O180 wall. 
Indeed, and in agreement to Refs.~\onlinecite{klomp_switching_2022,boddu_molecular_2017} we rather observe domain wall motion by  single-step local switching, nucleation and growth of clusters, see appendix.
Thereby, the dipole distribution along the perpendicular directions shows a broad distribution with and without domain walls or fields. 

As a second example, we pick R109 walls, which can be understood as O180 with additional homogeneous polarization along $x$. Here one may expect transient  $[11\bar{1}]$ and $[1\bar{1}1]$  domains after switching of one polarization component. In the R phase one may furthermore expect an even larger energy barrier to switch via the paraelectric state without polarization. 
Results for the R phase are collected in Fig.~\ref{fig:distR}. Indeed for a field of 300~kV/cm, we observe the nucleation and growth of $[11\bar{1}]$ and $[1\bar{1}1]$ domains on the wall with similar trends for the distribution of dipoles as for O180 walls.
We thus can reproduce the predictions by Landau-theory that Bloch-like R109 walls are possible and can be 
 interpreted as two coupled 71$^{\circ}$ walls.~\cite{taherinejad_bloch-type_2012} 
In agreement to the O phase, furthermore a fast bulk-like switching can be activated for larger fields cf.~Sec~\ref{sec:high}. Note that for the R phase, both the polarization and the applied field have a component along the wall normal and the high field switching can already be activated 
 for  300~kV/cm. In the shown example, the switching is dominated by domain wall motion. However the dipole-distribution in the first ps (not shown) hints to  monoclinic distortions with spacial correlation after the instantaneous field application. After 4~ps the system already relaxed back from this non-equilibrium state analogous to the discussion in Ref.~\onlinecite{khachaturyan_domain_2022}. However, the initial nucleation of $blue$ and $magenta$ clusters shows a corresponding spacial preference.

\subsection{High field switching}
\begin{figure}
 \includegraphics[width=0.35\textwidth]{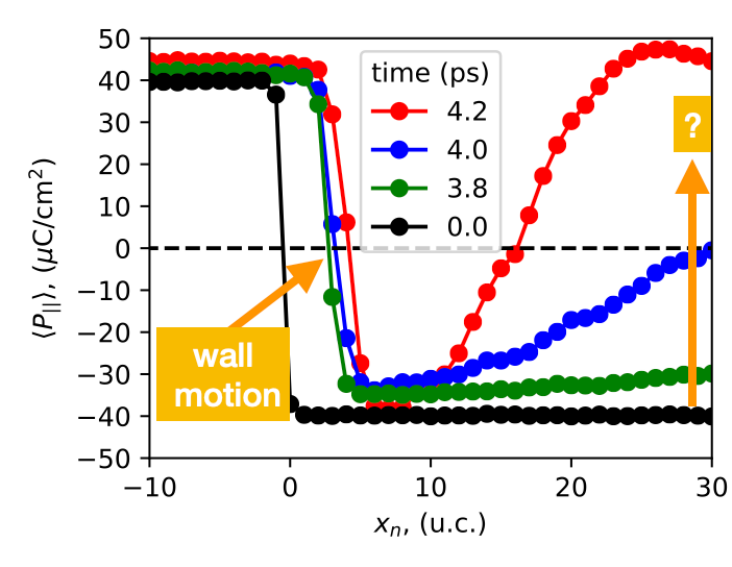} 
    \caption{Change of the polarization profile for high-field switching: While the O180 domain wall initially centered at $x=0$ moves through the system, a faster polarization rotation sets in in the center of the $black$ domain. 
}
    \label{fig:profile_high}
    \end{figure}

\label{sec:high}
\begin{figure}[t]
        \centering
        \includegraphics[width=0.5\textwidth,clip,trim=0cm 0cm 0cm 0cm]{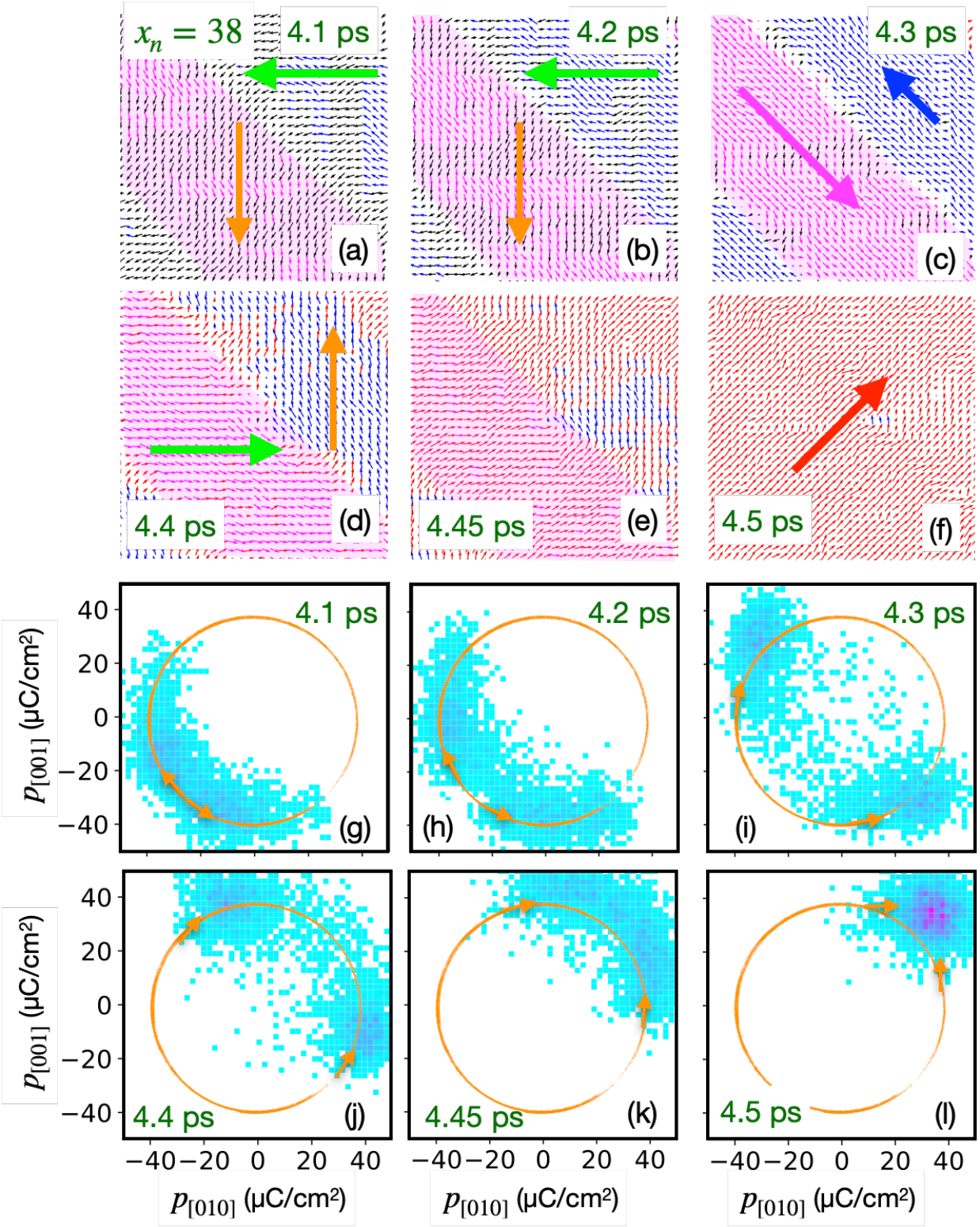}
        \caption{Time evolution of the dipole distribution in layer 38  for the bulk-like switching in field regime III   after
700~kV/cm have been applied along the $red$ direction
 (a)–(f) real space: Excerpt of the local dipoles colored via their classes. $magenta$ background marks onf of the newly formed 2d domains and large arrows highlight their macroscopic polarization direction. Blue: $[0\bar{1}1]$, Magenta: $[01\bar{1}]$, Orange: $\pm[001]$ and Green: $\pm[010]$
 (g)–(l) $p[010]--p[001]$ space: Colors encode the number
of dipoles per 2.06 $\mu$C/cm$^2 \times$ 2.06 $\mu$C/cm$^2$ and  the orange arrowheads highlight the continuous rotation
 path of the maxima of the distributions of both domains. 
        \label{fig:700_layer38}}
    \end{figure}
    In this section we discuss bulk-like switching in field regime III which can be 
     activated by high fields in the orthorhombic phase. As shown in Fig.~\ref{fig:profile_high} this switching is related to the rise of  $P_{||}$ in the center of the $black$ domain, e.g., for the given example layers with $P_{||}=0$ and positive saturation polarization exit well apart from the traveling initial domain wall after  about 4~ps and 4.2~ps, correspondingly. 
This is similar to the observed nucleation of a new domain apart from domain walls in high fields applied to the tetragonal phase.~\cite{boddu_molecular_2017}

Figures~\ref{fig:700_layer38}~(a)--(f) exemplifies the underlying microscopic processes in the O phase for one layer 
being initially in the center of the $black$ domain for an independent simulation run.
In contrast to the local nucleation and growth of $blue$ and $magenta$ clusters relevant for wall motion, the polarization on the whole slice breaks into two types of domains separated by equidistant walls  along $[0\bar{1}1]$.\footnote{We always find four domain walls in our simulation which is the minimal number of walls along $\langle 110\rangle$-type directions to fulfill periodic boundary conditions. }
In these domains, the polarization switches approximately uniformly by continuous polarization rotation.  The system shows a spatially correlated switching.
For instance, after 4.1~ps, most of the dipoles can still be classified as $black$, however, the polarization does not point along $[0\bar{1}\bar{1}]$ but rather scatters around $[0\bar{1}0]$ and $[00\bar{1}]$ in the  domains marked by green and orange arrows, respectively.
 Thus the layer is locally in a distorted T90 domain structure. Already after 4.3~ps, the local polarization rotates
 towards $\pm[0\bar{1}1]$ with O180 walls along $[0\bar{1}1]$ and finally the single domain $red$ state is reaches via an intermediate T90 configuration with $[010]$ and $[001]$ domains. 

Why does this homogeneous rotation of the polarization not induce large electric and elastic energy penalties?
First of all, there is no depolarization field, as both the intermediate T90 and the O180 domain structure are charge-neutral. Second, the domains are elastically compatible mimizing the elastic energy on the plane.\cite{marton_domain_2010} Third, the macroscopic strain relative to the surrounding $black$ matrix is minimized as polarization and strain rotation in both domains are fully symmetric. The latter can be clearly seen in the 
dipole distribution in $py-pz$ space shown in Figures~\ref{fig:700_layer38}~(g)--(l): After 4.1~ps two distinct maxima of the distribution have formed and rotate clock-wise and anti-clockwise towards the final single domain state. Thus a decrease of polarization and strain along $[010]$ in the orange domain is fully compensated by the corresponding increase of the green domain with equal area. Hover, as T90 walls are higher in energy than O90 walls, particularly in the temperature range of the O phase, this intermediate state can only be activated by high field strengths.

Excitingly the observed switching process in both domains is quasi-continuous, and thus in contrast to the commonly accepted picture of discrete ferroelectric switching. 
Although the dipole distribution shows clear maxima at each time step, these maxima rotate approximately continuously on the shown orange sphere and all intermediate states are populated, cf.\ the large weight of the distribution on the rotation path. 
For the intermediate O180 state (4.3~ps) furthermore some dipoles also populate the center of the plane however note that this does not correspond to direct 180$^\circ$ switching as the initial state is no longer existing in the excited material.
During the switching process, the complex 3D nucleus gradually expands towards the domain walls and once it reaches the moving wall, the whole switching is dominate by the 3D polarization rotation. 
The faster switching can be related to continuous 3D switching in a large volume fraction vs. discrete nucleation events followed by mainly 2D growth of nuclei for the domain wall motion.

\section{Conclusion}
We have used molecular dynamics simulations to study the underlying microscopic processes of field-induced switching of the orthorhombic phase of BaTiO$_3$. For moderate field strengths, we have shown that switching 
is dominated by two-dimensional nucleation and growth in front of travelling O180 domain walls. 
In contrast to the direct 180$^{\circ}$ switching on the well-studied tetragonal walls, the O180 walls thereby show a local realization of  multi-step 90$^{\circ}$ switching. In their transient state the moving walls split into 2 dimensional domains which are separated by one dimensional O90 walls and polarization vortices and host Ising-type and Bloch-type regions.

We have analyzed the time-evolution of this switching process and showed that the domain wall motion is dominated by thermally activated nucleation of clusters with 90$^{\circ}$ polarization relative to the applied field.
Size, width, probability of appearance of critical clusters, and  their growth rates increase with the field strength, this results in the expected increase of the domain wall velocity with field strength.
The final switching step follows different dynamics and is often related to the crossing of existing clusters rather than nucleation. 
It is thus promising to control the dynamics of the domain walls by means of tailoring the local energy barriers for 90$^{\circ}$ switching, for example, by defects,  inhomogeneities or straining.

We could relate the multistep switching by the low energy of the transient state on the O180 wall and the larger probability to switch one component of the polarization. These results can be transferred to other types of domain walls and a similar multi-step switching is possible for R109$^{\circ}$ walls.

For large fields, we have shown that a second ultra-fast multi-step switching channel can be activated. Thereby, homogeneous and uniform polarization rotation via an intermediate tetraongal 90$^{\circ}$ domain structure sets in in the middle of the domain and the 3 dimensional nucleus expands through the whole system within few picoseconds.
We thus observed a quasi-continuous switching process which may shed light on the understanding of microscopic dipole switching and boost the development of conceptually new switching models.

\section{Acknowledgement}
We acknowledge funding by the German research foundation (DFG) GR 4271/2.

\section{Data and code availability}
Scripts for the data analysis are openly available in the repository.~\cite{git} 
The library we used to analyze and characterize clusters is openly available in a git repository.~\cite{libferronucleation} Original data can be provided by the authors upon reasonable request.

\section*{Literature}
\bibliographystyle{apsrev}
\bibliography{mixed.bib}

\section*{Appendix}
In this appendix we collect additional figures.

\begin{figure}[t]
    \centering
    \includegraphics[width=0.4\textwidth,trim=0cm 0cm 0cm 0cm]{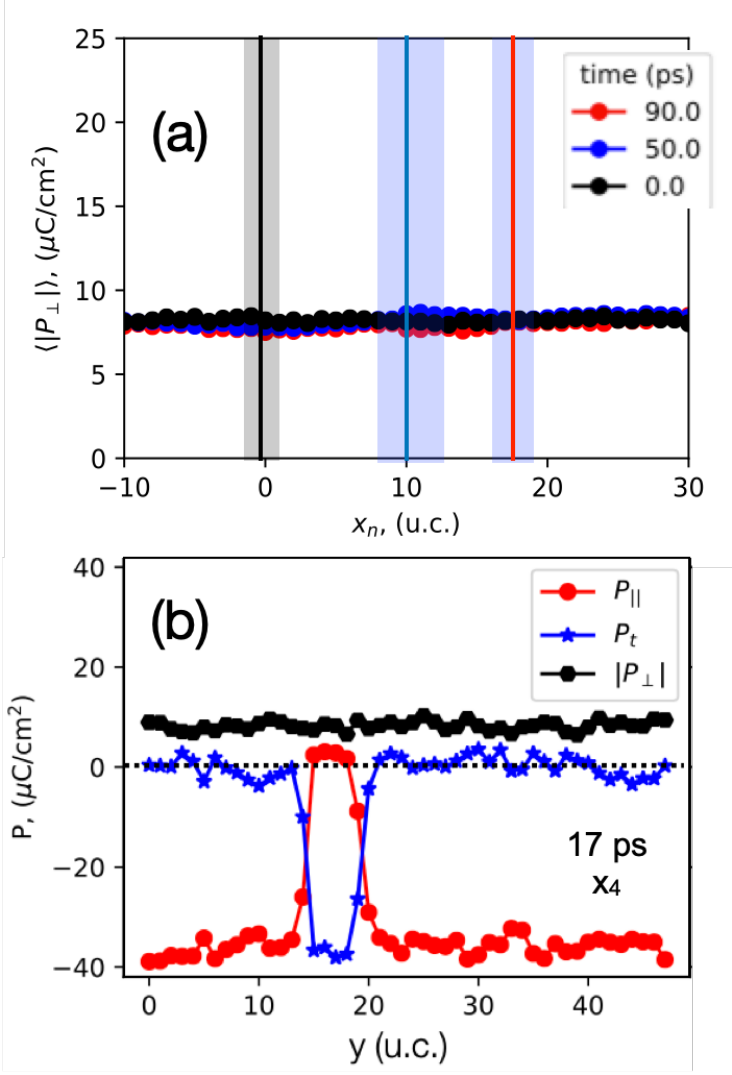}
    \caption{Change of the spatial distribution of dipoles with time after 500~kV/cm have been applied.  
    (a) Polarization profile  $P_{\perp}$, i.e.\ the polarization projection on [100], across the O180 domain wall 
    (black) and its time-evolution after an external field of  
   $E^{\text{ext}}=500$~kV/cm has been applied along [011]
   after 90~ps (red) 50~ps (blue) (cf.\ Fig.~\ref{fig:profile}). 
 Note that the sign of the projection has been removed before averaging. Vertical lines and shaded areas highlight the center and the width of the domain wall at the corresponding time step. 
   (b) Mean polarization on along [010] averaged over [001]-planes on $x_4$ shown in Fig.~\ref{fig:dist}. Indeed no N\'eel-like polarization component neither locally nor globally and sharp 90$^{\circ}$ domain wall.
 }
    \label{fig:cut_plane}
\end{figure}

 Figure~\ref{fig:cut_plane}~(a) shows the absence of Ne\'el polarization rotation across the moving orthorhombic domain walls at 0~ps (black), 50~ps (blue) and 90~ps (red) after an external field of 500~kV/cm has been applied for the same configuration shown in Fig.~\ref{fig:cut_plane}. Vertical lines mark the corresponding position of the domain wall centers.
In (b), all three polarization components $P_{||}$, $P_{t}$, and $P_{\perp}$ are shown across a one-dimensional domain wall centered at about 18~f.u. Also locally on that wall, no enhanced Ne\'el like polarization rotation exit.

\begin{figure}[t]
    \includegraphics[width=0.4\textwidth,clip,trim=0cm 0.2cm 0cm 0.8cm]{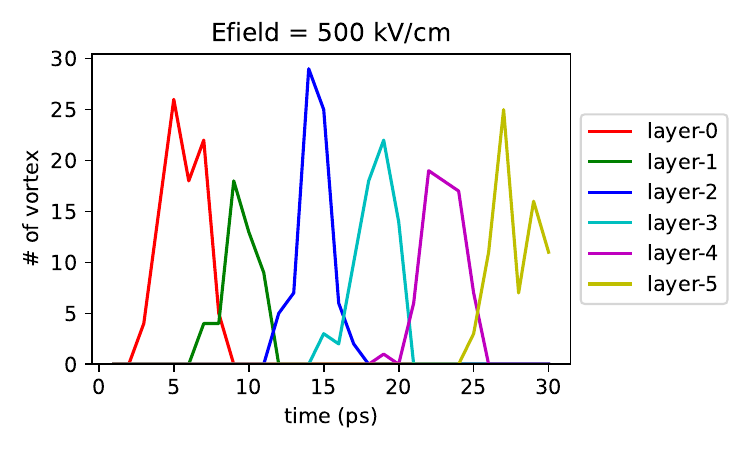}
        \caption{ Time-evolution of the number of vortices per on layers $x$ with an interface area of about 360 nm$^2$ after 500~kV/cm have been applied.
        \label{fig:vortices_50kv}}
    \end{figure}

Figure~\ref{fig:vortices_50kv} compares the time evolution of the number of vortices on different layers for the example of an external field of 500~kV/cm. For all layers this number is maximal if the wall crosses them.

\begin{figure}
 \includegraphics[width=.45\textwidth,clip,trim=0cm 0cm 0cm  0cm]{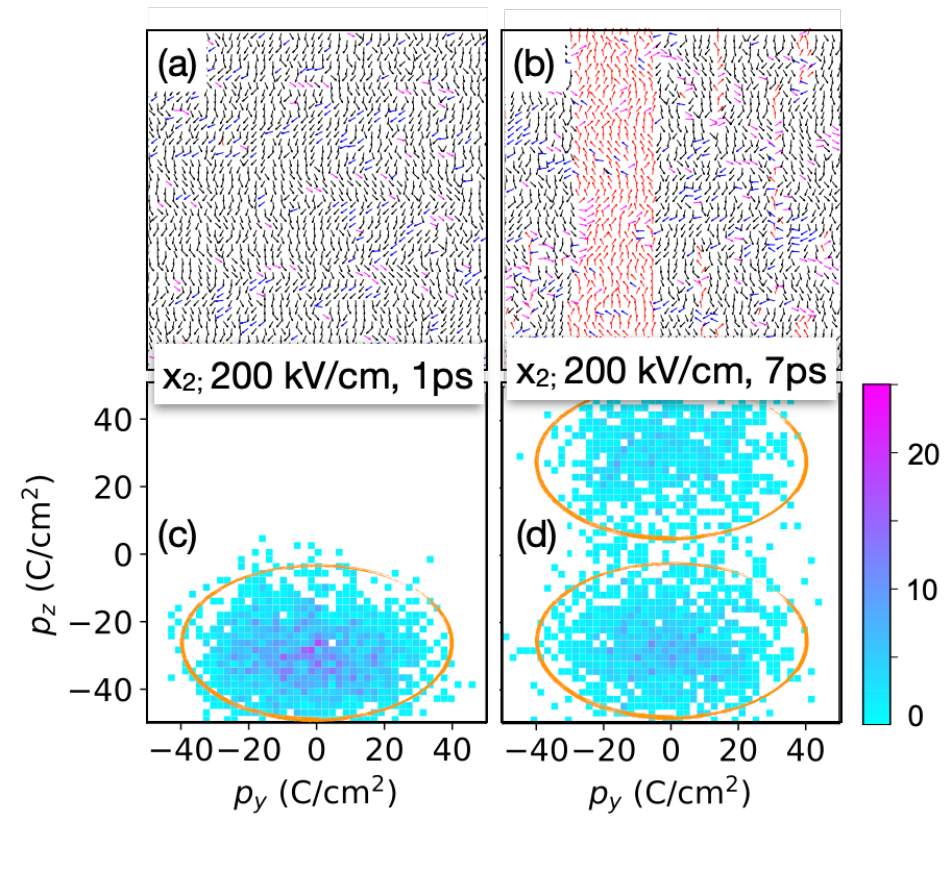}
        \caption{Change of the spatial  (a)-(b) and statistical (c)-(e) distribution of dipoles with time  (a),(c) 1~ps , (b),(e) 7~ps, after a field  has been applied along [001] to the tetragonal domain structure with [001] ($red$) and [00-1] ($black$) domains at 240~K. 
}
        \label{fig:distT}
    \end{figure}

Figure~\ref{fig:distT} shows the time evolution of the local polarization in the presence of a   moving tetragonal 180$^{\circ}$ wall. In subfigures (a)/(c) the domain wall has not reached the  the layer 2 and the dipoles scatter in a broad range of py-values around the tetragonal state with with finite $P_z$. In subfigures (b)/(d) the wall reached the plane, and dipoles in a 2 dimensional are have  switched by 180$^{\circ}$. The dipole distribution now scatters around both tetragonal states with $\pm P_z$

\begin{figure}
 \includegraphics[width=0.5\textwidth,clip,trim=0cm 0cm 0cm  0cm]{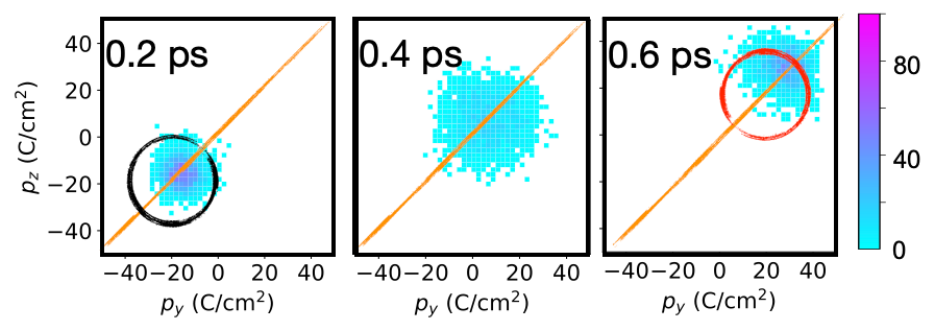}
        \caption{Single-step high-field switching of the rhombohedral phase (50~K): Change of the dipole distribution during field-induced wall motion (a)--(c) py-pz space on slice $x_n=30$ after an external field of 400~kV/cm has been applied along [111].
        Colors encode the number of dipoles per 2.06 $\mu$C/cm$^2$ 2.06 $\times$ $\mu$C/cm$^2$. Note that the distribution of $P_x$ is approximately constant with time. 
        }
        \label{fig:distR3}
    \end{figure}

In Figure~\ref{fig:distR3}, the time-evolution of the dipole distribution of the rhombohedral phase in the high field-regime is shown. The polarization switches quickly via the state with $Py=Pz=0$.

\end{document}